%% file: main.tex
\documentclass[aps, twocolumn, superscriptaddress, nofootinbib]{revtex4-1}
\usepackage{preamble}

\graphicspath{{./figures/}}

\begin{document}

\title{Mass-Spring Models for Passive Keyword Spotting: A Springtronics Approach} 
\author{Finn Bohte}
\author{Theophile Louvet}
\author{Vincent Maillou}
\author{Marc Serra Garcia}
\affiliation{AMOLF, Science Park 104, 1098 XG Amsterdam, the Netherlands}

\date{\today}

\input{tex/0_abstract}

\maketitle


\input{tex/1_introduction}

\input{tex/2_architecture}


\input{tex/3_msmodel}

\input{tex/4_performance}

\input{tex/5_conclusion}


\section*{Acknowledgements}
We thank Martin van Hecke for stimulating discussions.

Funded by the European Union. Views and opinions expressed are however those of the author(s) only and do not necessarily reflect those of the European Union or the European Research Council Executive Agency. Neither the European Union nor the granting authority can be held responsible for them.

This work is supported by the ERC grant 101040117 (INFOPASS).

\section*{}
Correspondence can be addressed to Marc Serra Garcia (m.serragarcia@amolf.nl).

\bibliography{ref}

\newpage
\clearpage


\end{document}

%% file: tex/0_abstract.tex
\begin{abstract}
Mechanical systems played a foundational role in computing history, and have regained interest due to their unique properties, such as low damping and the ability to process mechanical signals without transduction.
However, recent efforts have primarily focused on elementary computations, implemented in systems based on pre-defined reservoirs, or in periodic systems such as arrays of buckling beams.
Here, we numerically demonstrate a passive mechanical system---in the form of a nonlinear mass-spring model---that tackles a real-world benchmark for keyword spotting in speech signals. 
The model is organized in a hierarchical architecture combining feature extraction and continuous-time convolution,  with each individual stage tailored to the physics of the considered mass-spring systems.
For each step in the computation, a subsystem is designed by combining a small set of low-order polynomial potentials.
These potentials act as fundamental components that interconnect a network of masses.
In analogy to electronic circuit design, where complex functional circuits are constructed by combining basic components into hierarchical designs, we refer to this framework as springtronics. We introduce springtronic systems with hundreds of degrees of freedom, achieving speech classification accuracy comparable to existing sub-mW electronic systems.

\end{abstract}

%% file: tex/1_introduction.tex
\section{Introduction}
Computing is not limited to conventional electronic systems~\cite{jaeger2023toward}, but has been demonstrated in a  a variety of unconventional platforms, such as spins~\cite{wolf2006spintronics}, light~\cite{mcmahon2023physics}, and DNA~\cite{yang2024dna}.
Among these, mechanical systems stand out for three key advantages. 
First, they allow for direct processing of mechanical signals 
without transduction, as exemplified by passive sensors for spoken words~\cite{dubvcek2024sensor} or steps~\cite{barazani2020microfabricated}. 
Second, they can embody intelligent behavior within their structural dynamics, as demonstrated by agile soft robotic structures~\cite{sitti2021physical}. 
Third, they stand out for their low power dissipation, which enables efficient Internet of Things (IoT) devices~\cite{actuation2007multifunctional} and experiments on the fundamental energetics of information processing~\cite{dago2021information}. 
These advantages have sparked a renewed interest in mechanical computing, one of the earliest information processing platforms~\cite{bromley1982charles, freeth2006decoding}. 
Recent results emanating from this interest include mechanical logic gates~\cite{meng2021bistability}, an 8-bit processor~\cite{serra2019turing}, finite-state machines~\cite{liu2024controlled}, and reservoir computers~\cite{dion2018reservoir}.
However, prior work has primarily focused on elementary tasks such as counting~\cite{kwakernaak2023counting}, parity computation~\cite{coulombe2017computing}, and simple classification problems~\cite{dubvcek2024sensor}, where systems were predominantly designed with relatively simple architectures. 

In this work, we numerically demonstrate a mechanical system---in the form of a nonlinear mass-spring model---that passively implements spoken keyword spotting, a real-world signal processing task. Owing to its hierarchical, multistage architecture, the model rivals the accuracy of low-power electronic systems~\cite{cerutti2022sub} in a real-world, 12-class speech classification benchmark---representing a significant departure from the state-of-the-art in mechanical computing, which has been focused on single-stage systems and toy problems.


The platform of mass-spring models consists of networks of discrete masses connected through (nonlinear) springs.
These discrete models are established tools for studying complex mechanical phenomena, such as nonlinear effects in vibrations~\cite{raman1912experimental, duffing1918erzwungene}, allosteric responses in proteins~\cite{rocks2017designing} and locomotion in soft robots~\cite{urbain2017morphological}.
Importantly, discrete mass-spring models capture physical behavior independent of any specific realization.
For example, the same mass-spring model captures frequency conversion in both magnetic systems~\cite{serra2018tunable} and geometrically nonlinear structures~\cite{serra2016mechanical}.
This abstraction is a fundamental principle in electronic circuits design, where idealized components---resistors, inductors, capacitors, etc.---are combined in discrete circuit models before consideration is given to their physical implementation.
Beyond electronics, this discrete-model-first approach found recent success in condensed matter physics.
For instance, higher-order topological insulators were first discovered through discrete tight-binding models~\cite{benalcazar2017quantized} and later experimentally observed in physical systems~\cite{serra2018observation}.
Here, we introduce a similar, discrete-model-first, approach for mechanical computing, constructing models from a small set of idealized mass-spring components.
Inspired by circuit design, we refer to this approach as \emph{springtronics}.
We demonstrate sprintronics by designing a mass-spring model that embeds a speech classification architecture with analog feature extraction and a continuous-time convolutional neural network.

The present paper is organized as follows. In the remainder of the introduction, we establish the elements of the mass-spring models considered in this work.
In Sec.~\ref{sec:model architecture}, we describe the keyword spotting architecture, which is composed of signal processing operations that are compatible with these mass-spring models.
Section~\ref{sec:mass spring model} focuses on the implementation of this architecture as a springtronic model. The model is organized into subsystems, each mapping to one signal processing step from the architecture introduced in Sec.~\ref{sec:model architecture}. 
Next, Sec.~\ref{sec:performance} examines the speech classification accuracy and energetic efficiency of our mass-spring model. 
Finally, the paper concludes with a discussion and outlook in Sec.~\ref{sec:conclusion}.

\subsection{The space of mass-spring models}
\begin{figure}
    \centering
    \includegraphics{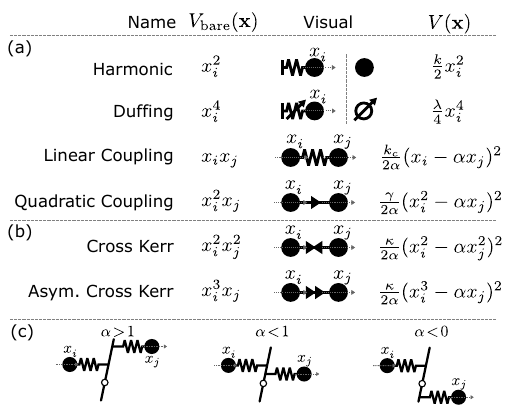}
    \caption{Visual depiction of springtronics elements: Masses are represented by circles and are connected through potentials. Dampings are typically not drawn, but can be depicted as a dashpot.
    Potentials ${V}(\ve{x})$ are positive definite and act on one or multiple masses, and are illustrated in the following ways:
    (a) Local harmonic potentials are depicted as springs connected to a support. For Duffing potentials, the spring is crossed by an arrow.
    For visual clarity, springs for local potentials can be omitted and the mass can be crossed by an arrow to indicate Duffing potentials.
    Linear coupling potentials are depicted as springs connecting two masses, and quadratic couplings are depicted as connecting lines overlaid with a triangle pointing towards $x_j$.
    Here, linearity refers to the force-displacement relation derived from the potentials.
    (b) The Cross Kerr coupling is depicted as a connection with two triangles pointing towards each other, and an asymmetric Cross Kerr coupling with two triangles pointing towards $x_j$.
    (c) Illustration of the leverage parameter $\alpha$. The coupling potentials are parameterized by a strength ($k, \gamma, \kappa$) and a leverage ($\alpha$). The leverage can be understood as the arm of a lever inserted between two segments of a spring that connects two masses.
    }
    \label{fig:springtronics building blocks}
\end{figure}

Throughout this paper, we use the term mass-spring model to refer to a network of masses interacting through force laws (Fig.~\ref{fig:springtronics building blocks}).
The mass-spring models are dynamic---with the masses subject to inertia and damping. The interactions considered here will all be conservative and reciprocal, with force laws derived from energy potentials acting on one or more degrees of freedom.
The degrees of freedom consist of the scalar-valued positions $x_i(t)$ and velocities $\dot{x}_i(t)$ of the masses. 
Even though we display the masses on the plane, they only represent one degree of freedom---and not two-dimensional in-plane displacements as in some works~\cite{rocks2017designing}. Moreover, the masses represent abstract degrees of freedom, such as structural modes or localized vibrations. 
Consequently, the model defines a topology rather than a geometry; 
the model dynamics are dictated by the connections between the masses rather than the location of masses, and are captured by equations of motion
\begin{equation}\label{eq:mass-spring-model}
    m_i \ddot{x}_i + b_i \dot{x}_i + \textstyle\sum_j\frac{\partial}{\partial{x_i}}V_j = f_i(t),
\end{equation}
where $i$ indexes the masses, $m_i$ is the mass, $b_i$ is the local damping, $V_j(\ve{x})$ for $\ve{x}=(x_1,\dots,x_n)$ are the potentials, and $f_i(t)$ is the external force.

Together with the masses and local dampings, interaction potentials form the elements of springtronics. We visually display them as shown in Fig.~\ref{fig:springtronics building blocks}.
The present work focusses on the four potentials listed in Fig.~\ref{fig:springtronics building blocks}(a), consisting of the monomials $V_{\mathrm{bare}}(\ve{x})=x_i^2$, $x_i^4$, $x_ix_j$, and $x_i^2x_j$.
These monomials form an expressive basis; complex functions, from digital logic~\cite{serra2019turing} to the speech recognition architecture implemented here, can be realized by combining these potentials. Moreover, they arise in a broad class of physical systems, because they constitute the low-order Taylor approximation of a generic nonlinear potential (the Taylor approximation, when truncated to fourth order in the energy, may also contain additional terms, such as those in Fig.~\ref{fig:springtronics building blocks}(b) that are not used in this work). 
The first two potentials from Fig.~\ref{fig:springtronics building blocks}(a) are local potentials, where $x_i^2$ yields a force linear to the displacement and $x_i^4$ yields a cubic force-displacement relation.
We refer to the latter as a Duffing term, but this same nonlinearity is known as the Kerr effect in optics.
The last two are coupling potentials, where $x_ix_j$ yields a linear coupling and $x_i^2x_j$ a nonlinear coupling. We refer to the linear coupling as a linear spring, and the nonlinear coupling as the quadratic coupling term. Physically, the latter interaction shows up in many systems in low-amplitude nonlinear regimes. In particular, it can be found in optomechanical cavities, for which rich intuition and experience has been accumulated~\cite{aspelmeyer2014cavity}. In mechanics, the coupling appears in systems such as guitar strings, where pulling longitudinally on one end increases the vibrational frequency due to geometric nonlinearity~\cite{raman1912experimental, conley2008nonlinear}.

A note of caution is required: Not all systems composed of the above elements have a lower bound on the potential energy.
This can lead to divergent dynamics. For example, consider a system with two degrees of freedom and potential energy landscape $x_1^2 - x_1^2x_2$, the system will experience divergence if the displacement $x_2$ exceeds $1$. These divergences are unphysical---they do not occur in actual experimental realizations---and indicate that the bare term does not exist independently. To ensure that the springtronic models only contain physically-meaningful interactions, we express our building blocks as combinations of terms $V_{\mathrm{bare}}(\ve{x})$ that respect positive definiteness. We do so by constructing positive-definite polynomials $V(\ve{x})$, that can be factorized into the square of a difference, and thus enforce a lower bound on the potential energy.
For the linear coupling $x_ix_j$, we take $x_i^2 - x_ix_j + x_j^2$ and write $V(x_i, x_j) = \frac{k_c}{2\alpha} (x_i - \alpha x_j)^2$, where $k_c$ denotes the $x_ix_j$ coupling strength, and $\alpha$ represents the leverage or arm of the coupling. The $\alpha$ parameterizes the relative contribution of the local terms towards positive-definiteness of $V$, and can be understood as the arm of a lever (Fig.~\ref{fig:springtronics building blocks}(c)). For typical springs, $\alpha = 1$, but in principle $\alpha$ can take any non-zero value.
For the quadratic coupling $x_i^2x_j$, we take $x_i^4 - x_i^2x_j + x_j^2$ and write $V(x_i, x_j) = \frac{\gamma}{2\alpha} (x_i^2 - \alpha x_j)^2$, where $\gamma$ denotes the $x_i^2x_j$ coupling strength, and $\alpha$ plays a similar role as in the linear coupling. Remarkably, the requirement for positive-definite potentials has measurable real-world consequences. For example, in the case of the quadratic coupling nonlinearity in a guitar string, positive-definiteness dictates that the boundary-induced frequency shift cannot exist independently from the longitudinal stiffness of the string, and from the Duffing nonlinearity of string vibrations.

\begin{figure*}
    \centering
    \includegraphics[width=180mm]{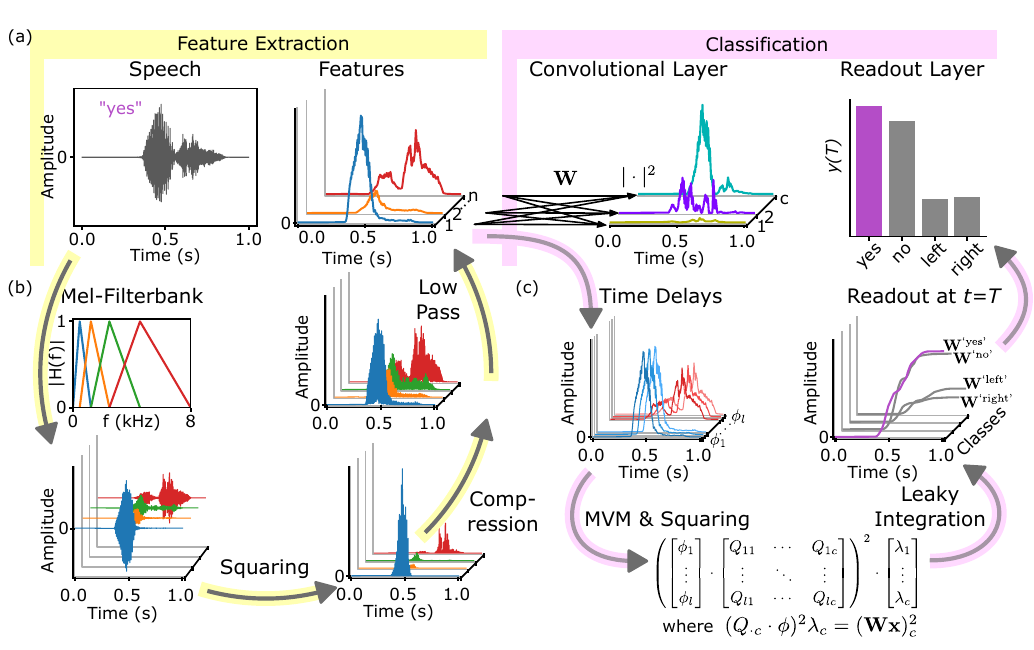}
    \caption{Model architecture: (a) The model consists of a feature extraction stage and a classification stage. The $n$ time dependent features capture spectral and temporal patterns of the speech signal. The extraction method is based on log-Mel spectrograms. Next, a convolutional classifier, composed of a CNN and a readout layer, predicts the likelihood of a keyword being present in the sound signal.
    (b) The features are extracted through signal filtering, signal squaring and signal compression operations. First, a Mel filterbank is applied, splitting the signal into multiple frequency bands. The resulting signals are squared, followed by a cubic root compression. Finally, a lowpass filter is applied to the signal.
    (c) The convolutional kernel weights $\mathbf{W}$ are realized via an instantaneous matrix-vector multiplication of time delayed copies of the features. We use a squaring activation function $x\mapsto x^2$ in the convolutional layer. The readout layer linearly combines the convolutional layer outputs, and performs a leaky integration over time. This integration yields the model readout. For each class, a different model is trained. The final prediction is given by the model producing the largest readout value.}
    \label{fig:KWS architecture}
\end{figure*}

With positive definiteness ensured, this work explores how to combine these potentials to create an artificial neural network. 
A challenge to overcome is that the building blocks of neural networks differ fundamentally from potentials above. 
Compared to neural networks, these mass-spring models exhibit reciprocal couplings, whereas neurons in a neural network break reciprocity. 
Additionally, the nonlinear interactions described above, and those typically arising in (micro-)mechanical systems, are generally captured by lower order polynomial terms contrary to common activation functions in artificial neural networks such as sigmoidal functions and rectified linear units. We will accommodate those differences by taking into account back-action when combining multiple stages, and by adapting the training process to work with the available nonlinearities. 

%% file: tex/2_architecture.tex
\section{Keyword Spotting Architecture}\label{sec:model architecture}
In this work, we design a mass-spring model for keyword spotting (KWS), a natural language processing problem where the task is to identify specific---sparsely occurring---keywords within speech signals.
It forms the basis, for instance, of wake word detection to activate voice assistants~\cite{lopez2021deep}.
Figure~\ref{fig:KWS architecture} provides an overview of the architecture of our system.
Here, we distinguish two stages in KWS systems: feature extraction and classification. 
For feature extraction, we base our architecture on log-Mel spectrogram features~\cite{davis1980comparison}, which capture temporal patterns in the signal, across predefined frequency bands.
In conventional digital implementations, these features are obtained by applying a Fourier transform to windowed segments of speech, squaring the magnitudes to compute the power spectrum, and filtering it with a Mel scale filterbank---triangular band-pass filters that are logarithmically spaced to mimic human auditory perception. 
Finally, logarithmic compression is applied to reduce the dynamic range, yielding the log-Mel spectrogram. 
The feature extraction stage of our system is composed of similar operations, but then tailored to the physics of the mass-spring models.
For classification, we base the model on a Convolutional Neural Network (CNN), which detects temporal and spectral correlations in the features, and is well-suited for KWS systems with limited computational resources.
Specifically, we consider a CNN architecture with temporal convolutions~\cite{choi2019temporal}.
The convolutions, parameterized by kernels, are followed by a nonlinear activation function.
The activation function in CNNs for speech typically is the rectified linear unit (ReLU). However, this activation function does not have a springtronic equivalent. Thus, we will replace it by a squaring activation function. The convolution in time will be implemented using a set of mechanical delay lines connected to a mechanical matrix-vector multiplication (MVM)~\cite{louvet2024reprogrammable}.
This section describes our KWS system in terms of analog, continuous-time, signal processing operations, and also discusses the training of the system. The mass-spring implementation will be discussed later in Sec.~\ref{sec:mass spring model}.

\subsection{Analog Continuous Time KWS Architecture}
This section introduces the analog, continuous-time KWS architecture that forms the basis of the mass-spring model. We preserve the operations from conventional digital KWS systems where feasible, replacing unsuitable blocks with alternatives that are more naturally adapted for mass-spring models. The feature extraction stage comprises three operations: signal filtering, signal squaring, and cubic root compression. 
First, a Mel filterbank splits the input into $n$ frequency-bin channels. Each filter output is then squared, rectifying the signal. 
We replace the logarithm with cubic root compression, which is more amenable to mass-spring implementation, while maintaining similar noise-reduction benefits~\cite{lyons2008effect}.
After compression, we apply a lowpass filter to remove high-frequency components introduced by the preceding squaring.
Unlike digital KWS systems, which partition signals into time windows consisting of tens of milliseconds, and then compute features via FFTs; analog systems process signals in continuous time. Still, we can achieve a similar temporal locality through the low-pass filter.

The analog classification model is composed of a single temporal convolutional layer with a quadratic activation function, succeeded by a linear readout layer.
Contrary to digital KWS, where convolutions are parameterized by discrete weights, continuous-time convolutions would typically require continuous densities, making optimization challenging. However, we implement a discrete parameterization for the continuous convolution, allowing it to retain the structure of its digital counterpart.
The convolutional layer is defined by a set of $c$ kernels with a perceptive field that spans all feature channels, and is localized at discrete points in the time domain.
The kernels can be formalized as weighted sums of Dirac measures---each defining a discrete measure.
The kernels are supported on a set of $m$ time points spaced $t_{d}$ apart, across all $n$ feature channels $x_i(t)$, and are parameterized by the weight tensor $\mathbf{W}\in\R^{n \times m \times c}$, or equivalently matrices $W^{(k)}\in\R^{n\times m}$ for $k = 1,\dots,c$. 
The $c$ convolution output channels are squared and passed to the readout layer, which consists of a leaky integrator acting over a sum across the channels.
This integrator is equivalent to a convolution with a kernel $e^{-{t}/{\tau}}$, parameterized by the integration timescale $\tau$.
The final model readout at time $T$ is given by
\begin{equation}\label{eq:conv-model}
    y(T) = \int_{0}^T e^{-\frac{(T-t)}{\tau}} \sum_{k=1}^{c} \left( \sum_{i=1}^n \sum_{j=1}^m W^{(k)}_{ij} x_i(t+jt_{d}) \right)^2 dt.
\end{equation}
For binary classification problems, the architecture predicts one word or the other depending on whether the readout $y(T)$ is above or below a threshold.
For multiclass classification, the architecture predicts each class using a one-versus-the-rest approach, training distinct models per class, and selecting the final prediction via an argmax (or softmax) over all readouts.~\cite{coulombe2017computing}

\subsection{Training}
CNNs are commonly trained with stochastic gradient descent. However, the model introduced in Eq.~\eqref{eq:conv-model} is a special case, since it can be rewritten as a support vector machine (SVM).
We take advantage of this property to efficiently train the classifier, and then transfer the weights to the corresponding mass-spring systems. For our models, the support vector machine is of relatively low dimension and will be trained using a squared hinge loss with $l_2$ regularization penalty. 
The model hyperparameters are the number of frequency bins $n$, the number of delays $m$, the delay time $t_{d} = 32$ ms, the read-out time constant $\tau = 80$ s, and the regularization strength for the SVM optimization. These values are common in speech processing and have not been extensively fine-tuned. The convolutional kernel shape is fixed by setting $c=nm$ so that $\mathbf{W}\in\R^{n \times m \times nm}$, resulting in $N = nm + \binom{n}{2}(2m-1)$ SVM features and coefficients.
In this work, we focus on small sized models, with few Mel filter bands $n=4, 6, 8, 12, 16$ and time delays $m=2,3,4$.
The features $\Phi$ used for SVM training are obtained by simulating the mass-spring model of the feature extraction stage (see Sec.~\ref{sec:mass spring model} for details), and numerically integrating the products $\phi_k^\top(t) \phi_l(t)$.  The SVM weights are then determined using a linear optimizer~\cite{fan2008liblinear}, that finds the coefficients through the primal optimization problem, since the number of features considered in this work will be less than the number of samples.

To determine the convolutional weights $\mathbf{W}$ from the SVM parameters, we start by deriving the SVM expression of the model from Eq.~\eqref{eq:conv-model}.
Let $\ve{\phi}(t)\in\R^{nm}$ denote the vector of the $n$ speech feature outputs at the $m$ discrete time points in the kernel support stacked at time $t$, specifically $\phi_l(t) = x_i(t+jt_{d}),$ for $l=j+(i-1)m$.
Now, taking $\lambda_k = |\sum_{i,j}W^{(k)}_{ij}|^{1/2}$ to be the norm of the kernel weight matrix and defining $Q_{kl} = \lambda_k^{-2}{W}^{(k)}_{ij}$, the integrand in \eqref{eq:conv-model} can be rewritten as
\begin{multline}\label{eq:classifier decompositio}
    \sum_{k=1}^{c} \left( \sum_{i=1}^n \sum_{j=1}^m W^{(k)}_{ij} x_i(t+jt_{d}) \right)^2  \\ = \sum_{k=1}^{c} \lambda_k \left( \sum_{l=1}^{nm} Q_{kl} \phi_l(t) \right)^2.
\end{multline}
In matrix notation, with $\Lambda = \operatorname{diag}(\ve{\lambda})$, one recognizes the quadratic form 
\begin{equation*}
    (Q^\top \phi(t))^\top \Lambda (Q^\top \phi(t)) = \phi^\top(t) C \phi(t), \text{ for $C =Q\Lambda Q^\top$.}
\end{equation*}
Substituting the above back in \eqref{eq:conv-model} and interchanging the summation and integration yields 
\begin{equation*}
\sum_{k=1}^{c} \sum_{l=1}^{nm} C_{kl} \int_{0}^T e^{-t/\tau}  \phi_k^\top(t) \phi_l(t) dt = \ve{c} \cdot \Phi,
\end{equation*}
where $\Phi \in \R^N$ forms the feature vector for the SVM, obtained by computing $\int_{0}^T e^{-t/\tau}  \phi_k^\top(t) \phi_l(t) dt $ for all possible pairwise combinations $(k,l)$ modulo symmetries in time delay and commutativity, and $\ve{c}$ is the unrolled matrix $C$ corresponding to the weights of the SVM.

%% file: tex/3_msmodel.tex
\section{Mass-spring model implementation}\label{sec:mass spring model}
\begin{figure}
    \centering
    \includegraphics{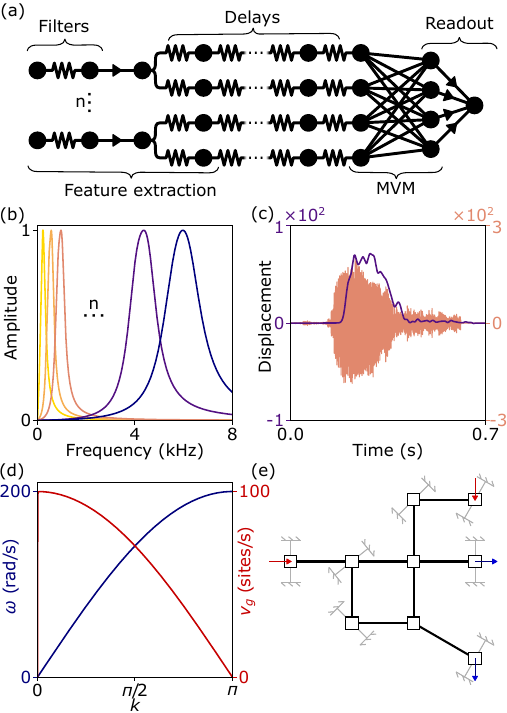}
    \caption{Mass-spring model overview:
    (a) Schematic representation of the speech classification mass-spring model. Subsystems are indicated with corresponding functionalities.
    (b) Amplitude response of the mass-spring approximation of the Mel filterbank. The mechanical filter responses are formed by combining two Lorentzians (as shown in Fig.~\ref{fig4}(b)). 
    (c) Example trajectory of a speech feature extracted by the mass-spring model (purple), with a frequency band-filtered speech signals obtained through the approximate Mel filterbank in (b) (orange).
    (d) Dispersion relation $\omega(k)$ for the delay line as per Eq.~\eqref{eq:dispersion_rel} (blue), with corresponding group velocity $v_g(k) = \frac{\partial \omega}{\partial k}(k)$ (red) for parameters $m=10^{-3}$, $k_c=10.0$, and $k_l = 10^{-5}$. 
    (e) Discretized representation of the geometry for MVM unit cell from \cite{louvet2024reprogrammable}. Note that the rectangles here represent masses that can move in-plane---corresponding to two springtronic degrees of freedom, and the straight lines represent geometric constraints. We construct an equivalent mass-spring model by assigning a large but finite stiffness to the geometric constraints, and then reducing the system to a set of input and output springtronic degrees of freedom. 
    }
    \label{fig3}
\end{figure}

This section describes how we design the mass-spring model that realizes the KWS architecture from Section~\ref{sec:model architecture}, by implementing the building blocks of the architecture as subsystems and interconnecting them.
An overview of the full mass-spring model is shown in Fig.~\ref{fig3}.
First, let us recall the building blocks of the speech classification architecture. The feature extraction stage consists of signal filtering, signal squaring, and signal compression. The convolutional layer of the classifier is composed of time delays, instantaneous matrix-vector multiplication, and squaring for the nonlinear activation function. Finally, the readout layer of the classifier combines the outputs of the convolutional layer and performs a leaky integration.
Here, we first discuss three building blocks that illustrate key design approaches for linear mass-spring models: signal filtering, time delays, and matrix-vector multiplications. Each of these functionalities is achieved following distinct approaches: optimizing eigenmodes, tuning dispersion curves, and utilizing zero-modes.
Next, we discuss how the matrix-vector multiplication is integrated with the delay lines to complete the convolution.
In the final subsection, we focus on the design of nonlinear building blocks by tuning the quadratic coupling. This subsection also covers the remaining steps required to complete and integrate the mass-spring speech classifier.

\subsection{Design of Linear Mass-Spring Models}
\subsubsection{Mechanical Mel Filterbank: Designing Frequency Responses through Eigenmode Engineering}
As discussed above, the first step in the feature extraction stage is signal filtering. Here, we present the design of a mass-spring system that acts as a frequency band-pass filter. The amplitude response is tuned to match triangular Mel filters by adjusting the system's eigenmodes.
The goal is to determine the parameters for a linear mass-spring model, with fixed topology, so that the system's frequency response $H(\omega)$ matches a predefined target response $T(\omega)$ in amplitude, while disregarding the phase of the response.
The frequency response of the system is obtained through diagonalization, allowing the system's dynamics to be decomposed into a set of uncoupled eigenmodes.
Combining the responses of the eigenmodes, the modal responses, yields the system response given by
\begin{equation*}
    H(\omega) = \sum_{k=0}^{n} \frac{F_k}{\omega_{0_k}^2  - \omega^2 + i \alpha \omega},
\end{equation*}
where $\omega_{0_k}$ denotes the modal frequency for eigenvalue $\lambda_k = \omega_{0_k}^2$, $F_k$ represents the external force coupling to each mode, and $\alpha$ is the modal damping coefficient under the assumption of mass-proportional damping (i.e. $b_i = \alpha m_i$ for each mass in Eq.~\eqref{eq:mass-spring-model}).
The optimization problem is to minimize the misfit $\|T(\omega) - H(\omega)\|^2$ over parameters $\omega_{0_k}$, $F_k$ and $\alpha$.

\begin{figure}
    \centering
    \includegraphics[width=\linewidth]{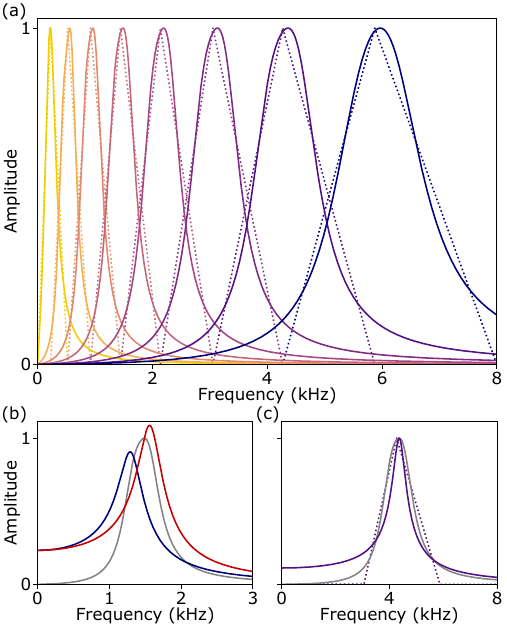}
    \caption{Mechanical Mel filters: (a) Amplitude responses of the optimized two-mode filterbank with 8 filters (full line) and the target Mel filter responses (dotted line). (b) Amplitude response of the 4th filter above in gray, with the responses of the two individual modes it is composed of: ${\alpha \omega_a^2}/{(\omega_a^2  - \omega^2 + i \alpha\omega)}$ (blue) and ${\alpha \omega_b^2}/{(\omega_b^2  - \omega^2 + i \alpha \omega)}$ (red), the difference of which defines the filter as per Eq.~\eqref{eq:filter-response}. (c) Amplitude response of the 7th filter above (gray) and an optimized single mode---Lorentzian---filter (purple), with the corresponding Mel filter (dotted line). The Lorentzian response is non-zero left of the Mel filter band, and the width of the peak deviates more from the width of the triangular response compared to the two-mode filter. These properties are related; the quality factor of peak-normalized Lorentizians determines both the response at $\omega=0$ and the width of the peak. The response of the filter has an impact on the classification accuracy. For the classification problem described in Sec.~\ref{sec:performance}, a two-mode filter achieves an accuracy of 81.4\%, compared to 75.2\% for a single-mode filter. As a reference, a digital Mel filterbank achives an accuracy of 82.0\%.
    }
    \label{fig4}
\end{figure}

Rather than resorting to a highly expressive system with many degrees of freedom or limiting ourselves to the simplest case of a single mode, we design the filters using two modes.
The key insight is that these modes can cancel out around $\omega = 0$ by coupling them with forces of opposing sign, as illustrated in Fig~\ref{fig4}(b). This results in a response profile that matches the shape of the target filter much more closely than a single mode, as shown in Fig~\ref{fig4}(c).
The transfer function of this system is given by
\begin{equation}\label{eq:filter-response}
    H(\omega) = 
    \frac{\alpha \omega_a^2}{\omega_a^2  - \omega^2 + i \alpha\omega}
    - \frac{\alpha \omega_b^2}{\omega_b^2  - \omega^2 + i \alpha \omega}, 
\end{equation}
and is parametrized by the frequency of the first mode $\omega_a$, the frequency of the second mode $\omega_b = \Delta \omega_a$, defined via a relative difference $\Delta$, and the damping parameter $\alpha$.
These parameters are optimized using gradient descent to minimize the misfit between Eq.~\eqref{eq:filter-response} and the triangular Mel filters, resulting in the filter responses depicted in Fig.~\ref{fig4}(a). For the optimization, we initialize $\omega_a$ at the center frequency of the Mel filter, set $\Delta = 1.1$, and choose $\alpha=400$.
We design a mass-spring system with this response using two linearly coupled masses, $x_a$ and $x_b$, each with the same local harmonic potential, as described by the following equations of motion:
\begin{align*}
    \ddot{x}_a + \alpha \dot{x}_a + k_{l} x_a  + k_{c} \left(x_a - x_b\right) &= f_a(t) , \\ 
    \ddot{x}_b + \alpha \dot{x}_b + k_{l} x_b  + k_{c} \left(x_b - x_a\right) &= f_b(t).
\end{align*}
By setting the masses to unity, the mass-spring parameters are derived from the modal parameters as follows: for the stiffnesses $k_l = \omega_a^2$, and $k_c = (\omega_b^2 - \omega_a^2)/2$, and for the force amplitudes $f_a \propto \alpha (\omega_a^2 +\omega_b^2),$ and $f_b \propto \alpha (\omega_a^2 -\omega_b^2)$ scaled to have peak response amplitude of $1$.

\subsubsection{Mechanical Delay Lines: Designing Time-Dependent Responses through Dispersion Engineering}
As discussed in Sec.~\ref{sec:model architecture}, the convolutional layer is designed via a matrix-vector multiplication on time-delayed copies of the extracted speech features. Here, we design the mass-spring model that induces these time delays. The model first splits each feature signal equally over $m$ chains of linearly coupled oscillators, which function as waveguides. We tune the propagation speeds of these waveguides to induce the relative delays.
Specifically, these chains will be referred to as delay lines, and the corresponding mass-spring model is described by
\begin{equation*}
    m \Ddot{x}_i + k_{c} \left(2 x_i - x_{i-1} - x_{i+1}\right) + k_{l} x_i = f_i(t),
\end{equation*}
where $m$ is the mass, $k_c$ is the coupling stiffness, $k_l$ is the local stiffness, and $i$ indexes the position along the chain. 
The wave propagation speed is determined by the group velocity, which follows from the dispersion relation, and is defined as the derivative of the angular frequency $\omega(k)$ with respect to the wave vector $k$.
For the delay line, the dispersion relation and the group velocity (in sites per second) are respectively given by~\cite{fetter2003theoretical}
\begin{align}\label{eq:dispersion_rel}
    \omega(k) &= \sqrt{\frac{k_l}{m} + \frac{4 k_c}{m} \sin^2{\left(\frac{k}{2}\right)}}, 
    &  \frac{\partial \omega}{\partial k} & = \frac{k_c}{m} \frac{\sin(k)}{\omega(k)}.
\end{align}
Notably, the group velocity is frequency dependent, so the delay line introduces signal distortion---different frequency components of the speech features propagate at different speeds.
We mitigate this distortion by minimizing the variation in $\frac{\partial \omega}{\partial k}$ over the relevant frequency range.
For a narrow frequency range, minimal variation is attained around the maximum of the group velocity, and this maximum can then be aligned with the frequency range of the signal.
However, for our speech feature signals, the frequency range was too broad, and we resorted to an alternative approach.
Instead, we simply operate the delay line close to the limit of $k_l \to 0$, by taking $k_l=10^{-6}k_c$.
This results in the group velocity being constant up to first order around $0$ frequency, where $\frac{\partial \omega}{\partial k} \approx \sqrt{m/k_c}$, as shown in Fig.~\ref{fig3}(d).
Subsequently, we obtain the desired time delay per site by tuning the mass $m$ and coupling stiffness $k_c$.
In the model, we first fix $k_c$ and then set the mass according to $m=t_d^2k_c,$ for time delay $t _d$.
For example, a $10\mathrm{ms}$ delay per site can be achieved by taking $k_c=10$ and $m=10^{-3}$.
Recall that the architecture described in Sec.~\ref{sec:model architecture} requires $m$ time delayed copies per feature. 
To achieve this, we simply connect $m$ delay lines to the features outputs wit the same propagation speed to preserve the mechanical impedance, and multiply the lengths.
In total, this leads to $nm$ delay lines, the outputs of which yield $\phi_l(t)$ from Eq.~\eqref{eq:classifier decompositio}.

\subsubsection{Matrix Vector Multiplication: Designing Instantaneous Linear Transformations through Zero-Mode engineering}
Recall that we construct the convolutional layer with the time delays above and an instantaneous matrix-vector multiplication that encodes the kernel weights of the convolutions.
Here, we adopt the mechanical system from \cite{louvet2024reprogrammable}, which proposed a geometry that performs an MVM under quasi-static conditions.
In our implementation, the mass-spring model uses idealized massless springs, eliminating dynamic effects.
For physical realization, the masses need to be small enough to avoid any internal resonances when the structure is operated under dynamic conditions.
Our MVM mass-spring model consists of a set of $nm$ input and $nm$ output masses, each of which is linearly coupled to all others, yielding a system with $2nm$ modes.
Among these, $nm$ are designed as zero-modes---deformation patterns along which the internal forces in the system remain balanced---so that they encode the matrix elements, or the convolutional kernel weights in our case.
Specifically, we encode the $nm \times nm$  unitary matrix $Q$ from Eq.~\eqref{eq:classifier decompositio}, which operates on the delay line outputs $\phi_l(t)$.
The zero-modes serve as the system’s effective degrees of freedom and have a theoretical modal stiffness of zero, while the remaining modes ideally exhibit infinite stiffness.
However, the stiffnesses of the nonzero modes in the mass-spring model inevitably remain finite.
By tuning the lowest nonzero mode frequency sufficiently above the delay line's operating band, we ensure that deformations along these modes remain negligible under the given input forces.

Finally, we complete the convolution by coupling the delay lines to the MVM to transmit the signals.
Efficient signal transmission between linear subsystems relies on impedance matching, which minimizes reflections at the coupling interface.
We achieve impedance matching by treating the MVM zero-modes as an extension of the delay line and coupling them with the same stiffness as between the delay line masses.
The zero-modes, in principle, have zero stiffness, so the tuning focuses on modal mass and damping.
The modal mass and damping are derived in the same way: Let us denote the $nm$ zero-modes of the system by  $\ve{q}_z$. Then, \[\begin{bmatrix}\ve{x}_{\mathrm{in}}\\\ve{x}_{\mathrm{out}}\end{bmatrix} = \begin{bmatrix}I\\Q\end{bmatrix} \ve{q}_z,\] since the zero-modes encode $Q$. In our model, the degrees of freedom of the MVM all have the same mass $m$ and damping $b$, leading to modal masses of $2m$ (and dampings of $2b$) because $Q$ is unitary:
\[\begin{bmatrix}I\\Q\end{bmatrix}^\top M\begin{bmatrix}I\\Q\end{bmatrix} = I^\top (mI) I+ Q^\top (mI) Q = 2mI.\]
Hence, we set the mass of the MVM input and output masses to half of those in the delay line, and the damping to half of the impedance-matched damping of the delay line.

\subsection{Design of Nonlinear Mass-Spring models}
\subsubsection{Nonlinear Operations with the Quadratic Coupling: The Paradox of Passive Squaring}
\begin{figure}
    \centering
    \includegraphics[width=\linewidth]{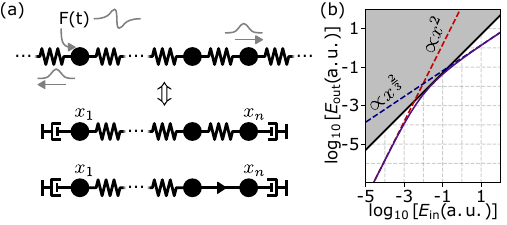}
    \caption{Energetics of the quadratic coupling: (a) The system consists of a truncated chain of linearly coupled masses, modeling an infinite delay line, with impedance-matched damping on both ends to simulate an infinite chain length. The system is excited from the left terminal site $x_1$. We study how the energy dissipated at the right terminal site $x_n$ changes when replacing the last linear coupling with a quadratic coupling. (b) Log-log plot of energy dissipated at $x_n$ ($E_{\text{out}}$) against energy input ($E_{\text{in}}$) for varying input pulse amplitudes, comparing the quadratic coupling (purple) to linear coupling (black), with reference lines for squaring (red) and power $2/3$ exponentiation (blue). The gray area is bounded by $E_{\text{out}} = \frac{1}{2} E_{\text{in}}$, corresponding to complete energy transfer where half of the input energy is dissipated at each terminal site, and indicates the area inaccessible due to energy conservation.
    }
    \label{fig:quadratic coupling energetics}
\end{figure}

In the remainder of this section, we will discuss how quadratic couplings can be used to design the nonlinear operations essential for both the feature extraction and the activation function of the CNN.
Recall that both of these stages rely on a signal squaring. Before moving to the design of the nonlinear mass-spring models, we first note that the squaring operation presents a paradox for passive systems: the operation increases signal energy for large amplitude signals.
Here, we will study the high-energy behavior of the quadratic coupling in a simple system to resolve this paradox.
We embed the quadratic coupling in an infinite delay line and excite it with Gaussian pulse signals of increasing amplitude, as shown in Fig.~\ref{fig:quadratic coupling energetics}(a).
The infinite delay line is approximated by truncating it and adding impedance-matched damping at the terminal sites.
We define the input energy as the work done by the input force $F(t)$ on the input site $x_1$, $E_{\text{in}} = \int F(t) \dot{x}_1(t) dt$.
The output energy is defined as the energy dissipated at the final site $x_n$, $E_{\text{out}} = \int b_n \dot{x}_n^2(t) dt$, where $b_n$ is the impedance matched damping value.
The results, shown in Fig.~\ref{fig:quadratic coupling energetics}(b), confirm that the quadratic coupling approximates squaring for low-energy input signals. 
For high energy input signals, the coupling approximates a power $3/2$ exponentiation.
This effectively achieves a cubic root compression combined with a squaring---revealing the motivation for the feature extraction method from Sec.~\ref{sec:model architecture}.

\subsubsection{Convolutional Activation Function and Leaky Integrator}
As discussed in Sec.~\ref{sec:model architecture}, the classification stage is composed of a convolutional layer and a readout layer. 
For the mass-spring model design of the convolutional layer, two steps remain: scaling the MVM output with $\lambda_k$ from Eq.~\eqref{eq:classifier decompositio}, and applying the squaring activation function.
We realize the squaring using a quadratic coupling, and encode the scaling in the strength of the coupling.
The quadratic coupling connects the convolutional layer to the readout layer in the mass-spring model.
The readout layer is implemented using a single mass that is coupled to all the MVM outputs.
First, we design the response of the readout mass for leaky integration.
This response is linear and parametrized by the frequency $\omega_r$ and quality factor $Q_r$.
The quality factor determines the decay of the response. In the overdamped regime, characterized by high damping relative to inertia (i.e. low quality factor $Q<0.5$), oscillatory systems exhibit exponential decay.
The rate of decay is given by the slow timescale of the system $\tau = 2Q_r/[\omega_r (1 - \sqrt{1 - 4Q_r^2})]$.
We fix $Q_r = 0.01$ and set $\omega_r \approx 1.25$ rad/s to match the timescale $\tau=80$ s of the readout kernel discussed in Sec.~\ref{sec:model architecture}.

\begin{figure}
    \centering
    \includegraphics{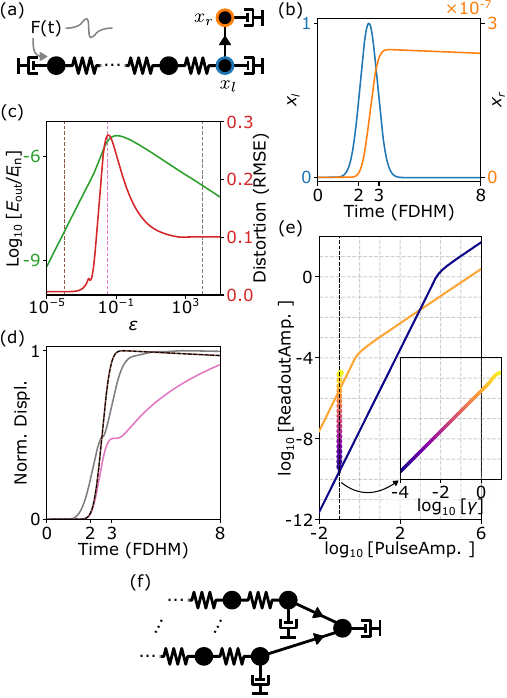}
    \caption{
    Designing a readout layer based on a leaky integrator: (a) A single-input, single output readout layer consists of a delay line excited by a pulse signal and terminated by impedance matched damping, where the terminal site $x_l$ (blue outline) is quadratically coupled to an overdamped mass $x_r$ (orange outline) corresponding to the integrating readout. 
    (b) The time-displacement curves of $x_l$ (blue) and $x_r$ (orange). The curve for $x_l$ shows the pulse signal that propagates over the delay line, and the curve for $x_r$ the leaky integrated pulse in the readout layer. 
    (c) Energy transfer and distortion against the squaring error parameter $\varepsilon$. Energy transfer is computed as the energy dissipated by the readout divided by the work done by the input force, and is shown in $\log_{10}$-scale. Distortion is computed as the root mean squared error between the readout trajectory compared to the leaky integrated square of the pulse signal (black, dashed line in (d)) each normalized to their peak.
    (d) Peak-normalized readout trajectories for $\varepsilon = \num{e-4}$ (brown), $\varepsilon = \num{e4}$ (pink), and high distortion $\varepsilon = \num{3.16e-2}$ (gray).
    (e) Tuning $\gamma$ to scale the readout. The readout displacement is plotted against input pulse amplitudes for $\gamma=\num{1}$ (orange) and $\gamma=\num{e-4}$ (blue). In the inset, the input pulse amplitude is fixed at \num{0.1} and $\gamma$ is varied, and the readout displacement for varying $\gamma$ is plotted (yellow to blue points) in log-log scale. The results show a linear relationship for $\gamma\leq\num{1}$ allowing for scaling of the readout through $\gamma$.
    (f) Extension of the system in (a) to the multiple input setting, which corresponds to the mass-spring model of Fig.~\ref{fig3}(a).
    }
    \label{fig6}
\end{figure}

Next, we turn to the quadratic coupling for the squaring activation function of the convolutional layer. 
In the mass-spring model for speech classification, each of the MVM outputs is connected to a single readout.
Here, we consider one quadratic coupling that connects a terminated delay line to a leaky integrator, as shown in Fig.~\ref{fig6}(a), and this setup straightforwardly extends to the multiple-input case illustrated in Fig.~\ref{fig6}(f). 
This system is a variation of the previous setup shown in Fig.~\ref{fig:quadratic coupling energetics}(a), but the quadratic coupling is now placed after the terminal site of the delay line that has impedance matched damping.
This placement reduces signal reflections from the quadratic coupling at the cost of a lower energy transfer---a trick we also use in the design of the classification mass-spring model.
To design the activation function, we are interested in the parameter regime of the quadratic coupling, where the system performs leaky integration of the squared input signal.
First, we analyze the equations of motion to find a nondimensional parameter associated with the squaring error.
Recall the quadratic coupling potential, and the resulting equations of motion as per Eq.~\eqref{eq:mass-spring-model}
\begin{align}\label{eq:motion quadratically coupled}
    m_l \ddot{x}_l + b_l \dot{x}_l + (k_l - 2\gamma x_r)x_l + 2\frac{\gamma}{\alpha} x_l^3 &= F(t), \\
    m_r \ddot{x}_r + b_r \dot{x}_r + (k_r + \gamma \alpha)x_r &= \gamma x_l^2,
\end{align}
where $F(t)$ is the input signal, $x_l$ corresponds to the delay line displacement, which is coupled to the readout mass with displacement $x_r$ (see Fig.~\ref{fig6}(a)). Illustrative time trajectories of $x_l$ and $x_r$ are shown in Fig.~\ref{fig6}(b).
Notice that the stiffness of $x_l$ has a cubic term and depends on the readout displacement.
We refer to the effect of this dependency as parametric back-action. 
Both the nonlinearity and the parametric back-action can cause errors in the squaring operation.
Next, we assume $k_r=0$, so the stiffness of $x_r$ stems only from the quadratic coupling via $\gamma\alpha$, which we tune through $\alpha$.
In turn, the resonance frequency of $x_r$ becomes $\omega_r = \sqrt{\gamma \alpha / m_r}$ and the frequency response of $x_r$ is given by
\begin{multline*}
    H_{x_r}(\omega) = \frac{\gamma F[x_l^2]}{\gamma \alpha -\omega^2 + i \omega b_r} = \frac{1}{\alpha} \frac{\omega_r F[x_l^2]}{\omega_r^2 -\omega^2 + i \omega (\omega_r/Q_r)}
\end{multline*}
which can be seen as $\alpha^{-1}$ times a linear operator acting on $x_l^2$.
Reintroducing this into Eq.~\eqref{eq:motion quadratically coupled}, we find that the parametric back-action becomes proportional to $\gamma/\alpha$, just as the nonlinear stiffness, making it the only parameter associated with the squaring error at a given input pulse amplitude.
The input pulse amplitude, that we will denote by $A$, is determined by the amplitude of the force and the stiffness scale of the delay line $k_l$.
Rescaling the displacement dimension for nondimensionalization of Eq.~\eqref{eq:motion quadratically coupled} reveals the error parameter depends on the square of the input pulse amplitude, yielding the squaring error parameter defined as $\varepsilon = A^2 \gamma\alpha^{-1}$. 
Figure~\ref{fig6}(c) shows the energy transfer and the signal distortion against $\varepsilon$. Readout trajectories for different values of $\varepsilon$ are shown in Fig.~\ref{fig6}(d).

Now that we understand the parametric regime of the quadratic coupling for squaring, we move to the scaling of the output.
Recall that this scaling implements multiplication of the $\lambda_k$'s from the convolutional kernels.
We continue with the assumption of $k_2=\num{0}$, which leaves the coupling strength $\gamma$ as the only remaining free parameter.
To examine the effect of $\gamma$, we perform a sweep over the input pulse amplitudes for two values for $\gamma$, while keeping $\gamma\alpha$ fixed by adjusting $\alpha$. The resulting readout amplitudes, shown in Fig.~\ref{fig6}(e), indicate a shift of the squaring regime.
We investigate how this shift scales with $\gamma$ by fixing the pulse amplitude at $\num{0.1}$ and varying $\gamma$, again for fixed $\gamma\alpha$. 
The inset in Fig.~\ref{fig6}(e) shows a linear relationship between $\gamma$ and the readout displacement. This relationship holds within the squaring regime, i.e. for small enough $\gamma$, and breaks down for large $\gamma$ when the squaring regime shifts beyond the amplitude of the fixed input pulse.
Bringing all of the above together, we tune the quadratic coupling by first fixing $\varepsilon$ based on $A$, which is derived from the extracted features. Then, $\gamma$ is determined by $\lambda_k$ of Eq.~\eqref{eq:classifier decompositio}, and $\alpha$ is set by fixing $\gamma\alpha$ to tune the integration time of the readout.

\subsubsection{Demodulation and Compression for Feature Extraction}
\begin{figure}
    \centering
    \includegraphics[width=\linewidth]{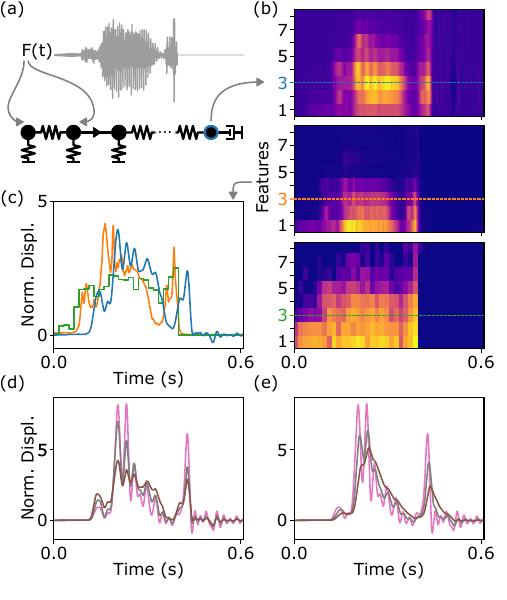}
    \caption{Mass-spring feature extraction:
    (a) Schematic representation of the feature extraction mass-spring model. The model consists of the two masses that act as the filter designed in Sec.~\ref{sec:mass spring model}. These masses are quadratically coupled to the low-pass filter, which is connected to a delay line with terminal damping. Take note that the lowpass mass has a local stiffness, contrary to the system in Fig.~\ref{fig6}(a), contradicting the assumption of $k_l=0$ used in the analysis of the latter. The speech sample that we excite the system with is shown in gray. The displacements plotted in panels (b)-(e) correspond to the right-most mass in the schematic (blue outline).
    (b)-(c) Comparison of spectrograms (b) and their third channel (c). Features are extracted via the mass-spring model (top, blue), direct evaluation of the architecture from Sec.~\ref{sec:model architecture} (middle, orange), and log-Mel spectrograms (bottom, green). Amplitudes are normalized for mean value.
    (d)-(e) Time-displacement curve of the third mass-spring feature at different parameters. In (d) for different values of the compression parameter $\alpha = 10^{-2}, 10^{-4}, 10^{-6}$ (pink, gray, brown) and in (e) for different lowpass cutoff frequencies of \qtylist{50;20;10}{\Hz} (pink, gray, brown). Displacement is normalized for mean value.
    }
    \label{fig:7}
\end{figure}

Recall from Sec.~\ref{sec:model architecture} that the feature extraction employs two nonlinear operations: signal squaring and cubic root compression.
In the mass-spring model, both operations are achieved concurrently through a quadratic coupling operated in the high energy regime.
This coupling connects each filterbank output to a mass whose linear response implement a low pass filter, completing our feature extraction stage.
In total, we construct the feature extraction subsystem using our mass-spring Mel filters, quadratically coupled to the low pass filter, which then fans out to delay lines, as illustrated in Fig.~\ref{fig:7}(a).
The features extracted using this mass-spring model approximate the features from Sec.~\ref{sec:model architecture} well and qualitatively resemble the log-Mel spectrogram, as shown in Fig.~\ref{fig:7}(b)-(c).

Finally, we design the remaining parameters for the feature extraction stage: the lowpass filter and the quadratic coupling.
We set the parameters of the lowpass filter by first fixing the local stiffness $k_{lp}$ and adjusting the mass $m_{lp}$ to match the specified cutoff frequency $f_c = \sqrt{m_{lp}/k_{lp}}$.
Recall that we fixed the low-pass frequency to \SI{50}{\hertz} in Sec.~\ref{sec:model architecture}. This value was motivated by visual inspection of the mass-spring features (see Fig.~\ref{fig:7}(d)) and assessing the model on the validation set of the task discussed in Sec.~\ref{sec:performance}.
The last parameter, the damping, follows from the quality factor of the filter via $b_{lp}=f_c/Q_{lp}$. In this work, we set $Q_{lp}=0.5$ for critical damping, which prevents resonance with minimal energy loss.
The damping of the mass has two origins: the local damping, and the impedance of the delay line which it is coupled to.
We account for this by setting the local damping to $b_{lp}$ minus the delay line impedance.
For the quadratic coupling, we fix $\gamma$ and tune $\alpha$ to control the compression. 
In this case, the parameter $\alpha$ determines in what regime of Fig.~\ref{fig:quadratic coupling energetics} we operate the coupling.
In terms of the potential, recall that the positive definite quadratic coupling introduces a nonlinear stiffness term $\frac{\gamma}{2\alpha} x_i^4$ on the filter output. 
Decreasing $\alpha$ lowers the compression threshold, leading to a more compressed feature signal.
The effects of compression strength, controlled through $\alpha$, and the lowpass cutoff frequency $f_c$ on the extracted speech features are displayed in Fig.~\ref{fig:7}(d) and (e), respectively.
Analyzing the behavior of the quadratic coupling for compression is more complex than for the activation function discussed previously, because we cannot assume $k_r=0$ for the system in Fig.~\ref{fig:7}(a), and we leave this for future work.

%% file: tex/4_performance.tex
\section{Classification performance and efficiency}\label{sec:performance}
In this section, we test our model using the Google Speech Commands Dataset (GSCD), and explore how the performance relates to the energy transfer of the system. 
The GSCD provides a widely used benchmark for KWS systems, especially for low-power devices and small-footprint models.
The dataset consists of \num{105829} \num{1}-second recordings of \num{35} words by \num{2618} speakers.
By default, 80\% of the data set is used for training, and the remaining is split evenly for validation and testing, both containing an equal number of samples for each class.
Additionally, background noise with a random gain is added to the recordings in the training set with a probability of \num{0.8}.
Before exciting the mass-spring models with the speech signal, we perform additional pre-processing of the data  that consist in centering the utterance in the time sequence, and subtracting signal components below the frequency range of speech. The latter is done by identifying slowly-varying signal background with a Savitzky-Golay filter ($f_c \approx \SI{75}{\hertz}$~\cite{schafer2011savitzky}) and subtracting it from the signal. 
The standard 12-class classification task is to distinguish a set of ten command words, ``Yes'', ``No'', ``Up'', ``Down'', ``Left'', ``Right'', ``On'', ``Off'', ``Stop'', and ``Go'', and two additional classes for ``Unknown Word'', comprised of the remaining \num{25} words, and ``Silence'' for recordings of background noise. The performance metric---the accuracy---is simply the percentage of samples for which the top predicted class matches the ground truth label. 
In addition to this standard GSCD test, we will also test the system on the four-digit binary classification task, also based on the GSCD, that was used in \cite{dubvcek2024sensor} on a mechanical system to compare it to prior work on mechanical speech recognition.

\subsection{Mass-Spring Model Classification Performance}
\begin{figure}
    \centering
    \includegraphics[width=\linewidth]{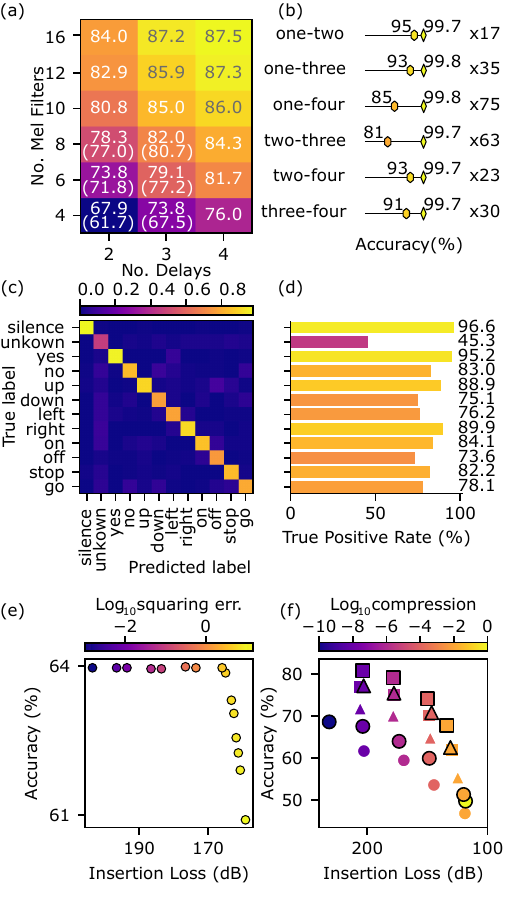}
    \caption{Numerical characterization of the speech classification system:
    (a) Test set classification accuracy for the GSCD 12-class task. The best found performance is shown for various model sizes of our KWS architecture evaluated digitally and implemented in mass-spring models (in brackets).
    (b) Binary classification accuracy for pairs of spoken digits ``one'' to ``four'' compared to the model from \cite{dubvcek2024sensor}.
    (c) Normalized confusion matrix with true positive rate for each class in (d) for the mass-spring model with 8 bins and 3 delays, yielding a classification accuracy of 80.7\%.
    (e) Accuracy vs insertion loss for varying squaring error $\varepsilon$ in the mass-spring CNN activation function. 
    (f) Accuracy vs insertion loss for different model sizes and feature compression levels. The mass-spring filterbank size is indicated by symbol (circle - 4, triangle - 6, square - 8) and the number of delays is indicated by outline color (white - 2, black - 3). Varying compression levels are shown in different fill colors. Squaring error is fixed at $\varepsilon = \num{1}$ for all models.
    }
    \label{fig:8}
\end{figure}

Here, we determine the speech classification performance of the mass-spring model designed in Sec.~\ref{sec:mass spring model} and discuss the results.
We simulate the system using a fixed time step 4th order Runge-Kutta algorithm~\cite{flannery1992numerical} implemented on a GPU using CUDA~\cite{sanders2010cuda}.
First, we excite the system with the speech files from the training set to compute the SVM features for training the model as discussed in Sec.~\ref{sec:model architecture}B.
The training results in a model for each class, which we then excite with the files from the test set. The predicted class follows from the model with the highest readout value.
The resulting test set accuracies are shown in Fig.~\ref{fig:8}(a) in brackets for various model sizes.
Our best-performing model achieved 80.7\% accuracy, with the corresponding confusion matrix shown in Fig.~\ref{fig:8}(c) and true positive rates for each class in Fig.~\ref{fig:8}(d).

To understand whether the observed classification error originates in the mass-spring implementation, or if it is a limitation of the chosen KWS architecture, we compare the classification accuracy with a digital realization of the signal processing pipeline from Sec.~\ref{sec:model architecture}. The evaluation results for various model sizes are shown in Fig.~\ref{fig:8}(a).
We found that the results are comparable to the mass-spring model; attaining an accuracy of 82.0\%, compared to the 80.7\% of the mass-spring model realization.
This indicates the mass-spring model approximates the targeted signal processing operations to a significant degree, underlining the success of the springtronics approach.
We suspect that a significant fraction of the mismatch can be attributed to the mass-spring realization of the Mel filterbank, as the difference decreases for higher number of filter bins.
Moreover, the digital evaluation allowed us to explore the performance of the architecture for larger model sizes, up to 16 Mel filter bins with 4 temporal delays per bin, which reached 88.1\% accuracy.
We hypothesize that significant room for improvement is left for the mass-spring model by both increasing the model size, and by tuning hyperparameters such as delay line time constants and filter parameters---potentially performing end-to-end backpropagation on the entire mass-spring model. 
However, these optimizations are left for future work.

Next, we compare our mass-spring model to electronic KWS systems.
For classification, small and efficient models have been developed for deployment on edge devices, which reach accuracies over 95\% (98.7\% for BC-ResNet~\cite{kim2021broadcasted}, 96.6\% for TC-ResNet~\cite{choi2019temporal}, and
98.8\% for WaveFormer~\cite{scherer2024tinyml}).
For these systems, the design commonly considers only the power consumption of the classification stage.
However, this constitutes only part of the energetic cost of complete systems, with typical signal acquisition, transduction and feature extraction taking 7.0 mW~\cite{cerutti2022sub} (for microphone, ADC and feature extraction).
At a sub-mW energy budget for the entire KWS system, Cerutti et al.~\cite{cerutti2022sub} reached 80\% in experiment,
compared to HelloEdge~\cite{zhang2017hello} which achieved 84.3\% while consuming of over 10mW. 
Since our mass-spring model integrates the entire KWS system and achieves 80\% accuracy, we conclude that it performs competitively with low-power electronic systems.

We also directly compared our mass-spring model against previous mechanical systems for speech classification and found a substantial increase in performance. 
The passive mechanical system introduced in \cite{dubvcek2024sensor} was primarily tested on a binary classification task of pairs of spoken digits ``One'', ``Two'', ``Three'', and ``Four''.
Specifically, the pair ``Two''-``Three'' posed a challenge for their linear model, plateauing at 59\% accuracy. A simple nonlinear mass-spring model was able to reach 81\% in this task. In contrast, the hierarchical nonlinear model proposed here achieves a 99.7\% accuracy for the pair ``Two''-``Three'', reducing the error by two orders of magnitude compared to prior linear work, and essentially saturating the benchmark.
The classification accuracy for each pair compared to \cite{dubvcek2024sensor} is shown in Fig.~\ref{fig:8}(b).
We also observe a reduction in classification error of over an order of magnitude for all other word pairs---highlighting the information processing capabilities of hierarchical mass-spring models designed using springtronics.

\subsection{Energetic Considerations}
In electronic KWS systems there is a trade-off between power consumption and classification accuracy, as studied in depth in \cite{cerutti2022sub}.
This subsection discusses the trade-off between the energy efficiency and accuracy in our model. 
Although the system is passive, energy is still dissipated during the signal processing. 
The loss of signal power between input and output, quantified as insertion loss, is primarily determined by dissipative losses and signal reflections.
These signal reflections can arise from impedance mismatches between linear systems or from nonlinearities such as the quadratic coupling.
Three aspects of the model influence both accuracy and efficiency: the model size, the parameters of the squaring activation function, and the feature extraction parameters.
Here, we numerically investigate their effects by training the model and evaluating its classification accuracy and insertion loss on the test set.

First, we fix the model size to a system with four frequency-bin filters and three time delays, and investigate the trade-off for the squaring and feature extraction.
For the squaring activation function, our goal is to balance the energy transfer and squaring error.
As shown in Fig.~\ref{fig6}(c), reducing the squaring error leads to a decrease in energy transfer.
To extend this analysis to the classification model, we examine how the squaring error parameter affects both the test set accuracy and the insertion loss. The results are depicted in Fig.~\ref{fig:8}(e).
We found the classification accuracy remains stable across a broad range for $\varepsilon < \num{1}$, but starts to degrade beyond this point.
This aligns with expectations from Fig.~\ref{fig6}(c) where the accuracy of signal squaring undergoes a similar transition.
To balance the trade-off for the squaring, we set $\varepsilon=\num{1}$. The performance reduction at lower insertion losses is likely tied to our training approach, which assumes perfect squaring.
A training strategy that accounts for the exact response of the quadratic coupling could extend its operational range, potentially reducing insertion loss further.

Next, for the feature extraction parameters, we explore the role of the design parameter $\alpha$. Recall from  Sec.~\ref{sec:mass spring model}B. that $\alpha$ controls the feature compression.
We observe that greater compression improves accuracy, but at the cost of higher insertion loss, until a saturation is reached around $\alpha<\num{e-10}$. Conversely, the reduction in insertion loss for lower compression saturated around $\alpha>\num{1}$. 
Regarding model size, Fig.~\ref{fig:8}(f) shows that increasing the model size improves accuracy with minimal impact on insertion loss, particularly at higher compression.
Beyond the aspects of the model discussed here, we suspect that optimizing unexplored parameters---such as the integration timescale of the readout layer---could further reduce insertion loss.
Another factor limiting efficiency is the reliance on the squaring activation function. Relaxing the assumption of perfect squaring and instead leveraging the exact dynamics of the quadratic coupling may lead to a more efficient convolutional layer.

%% file: tex/5_conclusion.tex
\section{Conclusion and outlook}\label{sec:conclusion}
In this work, we have demonstrated a passive mass-spring model for speech classification, achieving accuracy comparable to low-power electronic systems.
While nonlinear mass-spring models have previously been explored for information processing, existing approaches have relied on reservoir computing~\cite{coulombe2017computing}---where the system is parametrized randomly---or on repeating logic gates~\cite{serra2019turing}. 
In contrast, we designed a modular, hierarchical mechanical system---consisting of filters, demodulators, delay lines, matrix-vector multiplications, activation functions and leaky integrators, each performing a relevant step of the computation. These sub-systems are constructed from a small set of discrete, idealized components. Drawing an analogy to electronic circuit design, we refer to this design approach as springtronics.
By achieving competitive performance with low-power electronic solutions, our mass-spring models demonstrate that relatively simple systems---rooted in widely accessible classical mechanics---can perform complex information processing tasks. These systems are thus an elegant framework for theoretical studies on the physics of computation---where abstract information processing tasks must be embodied on a concrete physical realization to investigate their characteristics---and for potential applications on low-power MEMS devices.

This study also suggests several future work directions. In particular, realizing the device experimentally requires translating the mass-spring models into structural geometries, e.g., by leveraging methods such as~\cite{vanel2017asymptotic, matlack2018designing}. Since springtronics is a generic information processing platform, future works should also explore which other tasks are suited to mass-spring computational building blocks (e.g., structural health monitoring, step counting). Finally, the proposed design can be further optimized, both through architectural refinements and through end-to-end training, to improve the efficiency-accuracy Pareto front.
The structured design approach for mass-spring models introduced in this work may contribute to the realization of efficient, low-power mechanical computing for various tasks.